\documentclass[useAMS,usenatbib,referee]{biom}

\def\bSig\mathbf{\Sigma}

\input{anc/DW_Macros}
\usepackage{rotating}
\usepackage{amsmath,amssymb}
\usepackage{arydshln}
\usepackage{float}
\usepackage{multirow}
\usepackage{soul}
\usepackage{multicol}
\usepackage{booktabs}
\usepackage{algorithm}
\usepackage{algpseudocode}
\usepackage{graphicx}
\usepackage{xcolor}
\title[A group testing based exploration of age-varying factors in chlamydia infections among Iowa residents]{A group testing based exploration of age-varying factors in chlamydia infections among Iowa residents}

\author{Yizeng Li,
Dewei Wang$^{*}$\email{deweiwang@stat.sc.edu}, and Joshua M. Tebbs \\
Department of Statistics, University of South Carolina, Columbia, South Carolina, U.S.A.}

\begin{document}

%  This will produce the submission and review information that appears
%  right after the reference section.  Of course, it will be unknown when
%  you submit your paper, so you can either leave this out or put in 
%  sample dates (these will have no effect on the fate of your paper in the
%  review process!)

% \date{{\it Received October} 0000. {\it Revised February} 0000.  {\it
% Accepted March} 0000.}

%  These options will count the number of pages and provide volume
%  and date information in the upper left hand corner of the top of the 
%  first page as in published papers.  The \pagerange command will only
%  work if you place the command \label{firstpage} near the beginning
%  of the document and \label{lastpage} at the end of the document, as we
%  have done in this template.

%  Again, putting a volume number and date is for your own amusement and
%  has no bearing on what actually happens to your paper!  

\pagerange{\pageref{firstpage}--\pageref{lastpage}} 
\volume{Manuscript}
\pubyear{2024}
\artmonth{December}

%  The \doi command is where the DOI for your paper would be placed should it
%  be published.  Again, if you make one up and stick it here, it means 
%  nothing!

\doi{}

%  This label and the label ``lastpage'' are used by the \pagerange
%  command above to give the page range for the article.  You may have 
%  to process the document twice to get this to match up with what you 
%  expect.  When using the referee option, this will not count the pages
%  with tables and figures.  

\label{firstpage}

%  put the summary for your paper here

\begin{abstract}
Group testing, a method that screens subjects in pooled samples rather than individually, has been employed as a cost-effective strategy for chlamydia screening among Iowa residents. In efforts to deepen our understanding of chlamydia epidemiology in Iowa, several group testing regression models have been proposed. Different than previous approaches, we expand upon the varying coefficient model to capture potential age-varying associations with chlamydia infection risk. In general, our model operates within a Bayesian framework, allowing regression associations to vary with a covariate of key interest. We employ a stochastic search variable selection process for regularization in estimation. Additionally, our model can integrate random effects to consider potential geographical factors and estimate unknown assay accuracy probabilities. The performance of our model is assessed through comprehensive simulation studies. Upon application to the Iowa group testing dataset, we reveal a significant age-varying racial disparity in chlamydia infections. We believe this discovery has the potential to inform the enhancement of interventions and prevention strategies, leading to more effective chlamydia control and management, thereby promoting health equity across all populations.

%The results showcase the effectiveness of our approach in handling the complexities of group testing data and providing reliable inference for both assay accuracy and valuable insights into disease prevalence dynamics.

%In the study of chlamydia, understanding racial disparity is essential for devising targeted interventions, improving healthcare access and quality, addressing social determinants of health, and ultimately achieving health equity for all populations.
%Group testing, which involves testing subjects in pooled samples rather than individually, is a popular cost-saving strategy for screening rare diseases. Statistical analysis of group testing data faces challenges brought by the concealed individual disease status due to pooling and imperfect testing. By adopting the Bayesian framework to address these challenges, this article proposes a new varying coefficient mixed model with variable selection. The model incorporates a random effect to account for potential geographical factors, allows regression associations to vary with a covariate like age, and employs variable selection to identify significant covariates and trends that vary with age. In addition, our method can estimate all the unknown assay accuracy probabilities. The performance of our model is accessed by comprehensive simulation studies and a chlamydia group testing dataset. The results showcase the effectiveness of our approach in handling the complexities of group testing data and providing reliable inference for both assay accuracy and valuable insights into disease prevalence dynamics.
\vspace{.1in}
% \vspace{1in}
% The varying coefficients can depict 

% This is made possible by permitting regression coefficients to vary with age, which we effectively model using Gaussian predictive process priors. In addition, our method can simultaneously estimate unknown assay accuracy probabilities and accommodate potential geographical factors through random effects. However, our framework treats each regression coefficient as an unknown function of age, potentially resulting in an overparameterized model. Therefore, we suggest a stochastic search variable selection procedure to automatically identify fixed and age-varying effects, which popularizes the use of the spike-and-slab prior in generalized linear regression. To evaluate the performance of our model, we conduct comprehensive simulation studies and apply the proposed method to group testing data related to chlamydia infection in Iowa. The results clearly demonstrate the effectiveness of our approach in handling the complexities of group testing and providing reliable estimates for both assay accuracy and regression coefficients. This research contributes significantly to the advancement of group testing methodologies, offering a robust statistical framework that can handle intricate data structures and provide valuable insights into disease prevalence dynamics across different age groups and predictors.
\end{abstract}

%  Please place your key words in alphabetical order, separated
%  by semicolons, with the first letter of the first word capitalized,
%  and a period at the end of the list.
%

%  As usual, the \maketitle command creates the title and author/affiliations
%  display 

\maketitle

\section{Introduction}
\label{sec1}

Chlamydia, commonly causing cervicitis in females \citep{workowski2013Chlamydia}, is one of the most frequently reported bacterial sexually transmitted diseases (STDs) in the United States. Untreated chlamydia can cause pelvic inflammatory disease, potentially leading to ectopic pregnancy, chronic pelvic pain, and infertility \citep{land2010epidemiology}. Though there are effective treatments, many infected women might not seek them because chlamydia often has no symptoms while attacking their reproductive system. Acknowledging the asymptomatic threat, the Centers for Disease Control and Prevention (CDC) advises regular chlamydia testing for sexually active women. In many states, such as Iowa, the government offers assistance to provide no-cost or affordable testing for chlamydia and other STDs. Nevertheless, the financial burden on testing agencies becomes significant when the number of testees is large, prompting the need to explore cost-effective solutions \citep{roberts2006screening}.

As part of Iowa’s surveillance program for chlamydia infection, the State Hygienic Laboratory (SHL) has employed group testing as a solution. Its testing protocol, initially introduced by \cite{dorfman1943detection} to screen World War II recruits for syphilis, involves assigning arriving specimens at SHL into different groups. Within each group, the specimens are mixed to form a master pool, which is then tested. If it tests negative, patients contributing to that group are declared negative; otherwise, they undergo separate testing for a final diagnosis. Consequently, patients in a negative master pool are diagnosed with just one test. For low-prevalence diseases like chlamydia, group testing can save costs substantially. According to \cite{tebbs2013two}, the SHL has realized an annual savings of approximately \$0.62 million in chlamydia screening for Iowa residents.

In addition to all testing outcomes, the SHL dataset also encompasses various individual-level information from each patient, including age, race, recent sexual behaviors, presence of symptoms, the clinic site where the specimen was collected, etc. This unique dataset has established itself as an invaluable resource for chlamydia studies. %and has significantly advanced the development of statistical regression in group testing.

Analyzing the SHL dataset poses a challenge due to the inherent complexity of its group testing data structure. Early group testing research focused on estimating the disease prevalence without considering covariates \citep[see][for a review]{liu2012optimality}. Subsequent regression studies initially relied on outcomes of master pool testing \citep{vansteelandt2000regression,bilder2009bias, huang2009improved, delaigle2011nonparametric, wang2013semi, delaigle2014new}, but these methods could not incorporate information obtained from retesting patients in positive master pools at SHL. More appropriate methods, which include but are not limited to recent parametric approaches \citep{xie2001regression,mcmahan2017bayesian} and semiparametric regression techniques \citep{wang2014semiparametric,liu2021generalized}, have since emerged. The key distinction lies in the fact that parametric approaches enforce a linear relationship between covariates and the log-odds of disease risk. Conversely, semiparametric methods offer flexibility, allowing for the identification of non-linear covariate effects. For example, \cite{liu2021generalized} applied the generalized partially linear additive model to the SHL dataset, employing a non-linear function to capture the age effect while controlling other covariate effects as linear. Their non-linear estimate revealed a peak infection risk occurring around the age of 18, with a noticeable rise in risk for females aged 50 and above, a finding that aligns more closely with the current understanding of chlamydia infections compared to treating the age effect as linear.

% Alongside age, racial disparity also plays a significant role in understanding and addressing the epidemiology of chlamydia infections.
Alongside age, racial disparity also plays a significant role in understanding and addressing the epidemiology of chlamydia infections. In the existing research on the SHL dataset, the association between race and infection risk has been assumed to be independent of age. However, other chlamydia research has shown that this assumption might not be true. For example, a recent investigation focusing on women aged 15–34 years individually tested in Washington \citep{chambers2018racial} revealed a notable racial disparity trend in the cumulative risk of chlamydia diagnosis. The cumulative risk exhibited a slower increase in non-Hispanic whites compared to other races before age 25. After reaching 25, the rate became similar across all racial groups. This intriguing observation suggests that the racial disparity in chlamydia infections might vary with age. If this holds in Iowa as well, researchers and policymakers could develop more targeted interventions and prevention strategies to enhance the control and management of chlamydia. Unfortunately, the current group testing regression methods are unable to investigate age-varying associations.

We provide a remedy in this article by using varying coefficient models. These models, initially proposed by \cite{cleveland1992statistical} and \cite{hastie1993varying}, allow the regression coefficients to vary with a chosen covariate (in our case, age) and have proven to be a powerful tool in the semiparametric regression toolbox. However, they have never been extended to group testing, and we aim to make the first attempt. 

Our approach is developed within the Bayesian framework to flexibly accommodate different group testing strategies and estimate the unknown testing accuracies (i.e., test sensitivity and specificity). Gaussian predictive process priors (GPPs) \citep{banerjee2008gaussian} are used to estimate all the varying coefficients nonparametrically. A notable challenge for varying coefficient models in group testing arises from heavily imbalanced data due to the rarity of the disease. This imbalance can result in wide confidence intervals and non-informative inference. Therefore, it is important to regularize the estimation. Herein, we propose regularizing our estimation using a stochastic search variable selection (SSVS) process. This process categorizes each covariate into one of three groups: (i) \textit{insignificant covariates}; (ii) \textit{significant but age-independent covariates}; (iii) \textit{significant and age-varying covariates}. Our regularization differs from those used for selecting linear covariate effects in group testing, as seen in the works of by \cite{gregory2019adaptive}, \cite{lin2019regression}, and \cite{joyner2020mixed}.

Another aspect we must consider is that the SHL specimens were collected from a wide range of clinic sites throughout the state, such as primary care, community health, and women's health clinics, for centralized testing and analysis. The inherent disparities among rural, urban, and suburban areas, as well as different clinic types, as outlined in \cite{bergquist2019presenting}, underscore the need to address heterogeneity across subgroups at different locations. Using random effects, \cite{chen2009group} and \cite{joyner2020mixed} have confirmed this source of heterogeneity. We follow them in developing our varying coefficient model for the SHL group testing data. To facilitate efficient posterior inference, we have developed a Markov chain Monte Carlo (MCMC) sampling algorithm that integrates GPPs, SSVS, random effects, and the estimation of unknown testing accuracies. While our study is motivated by the SHL data, the method we propose applies to data from all sorts of group testing algorithms, whether or not random effects need to be considered.

The remainder of this article is organized as follows. Section \ref{sec2} provides preliminaries for the proposed varying coefficient mixed model with variable selection as well as model assumptions. Section \ref{sec3} details the data augmentation procedures and prior elicitation. Section \ref{sec4} presents the posterior sampling algorithm steps. Section \ref{sec5} evaluates the effectiveness of our methodologies through a comprehensive simulation study. In Section \ref{sec6}, we conduct an in-depth analysis of the SHL group testing data. Section \ref{sec7} concludes this article with a summary and future research prospects. Technical details about posterior sampling steps and their derivations along with additional numerical results are provided in the Web Appendices.

% ----------------------------------------------------------------------------------------------------------------

\section{Methodology}
\label{sec2}
\subsection{Notations and the model}\label{sec-notation-preliminaries}
Suppose we have $N$ individuals undergo screening for an infectious disease. Each of the $N$ individuals visits one of $L$ distinct clinics over the state. The clinics collect specimens
%since we have specimens here.
(e.g., blood, urine, or swabs) from the individuals and subsequently ship these collected samples to a laboratory (such as the SHL) for testing. A group testing algorithm is followed in the laboratory to screen all the specimens.

With the group testing data, our objective is to estimate an underlying individual-level model. To elucidate the model, we use $\tlY_i$ to represent the true infection status of the $i$th individual, where $i=1,\ldots, N$. Here, $\tlY_i=1$ (or $0$) indicates that the participant is \textit{truly} positive (or negative). Additionally, we denote the age of the $i$th participant as $u_i$ and other $p$ covariates as $\blx_i=(x_{i1},\ldots,x_{ip})^\t$. We consider the following varying-coefficient mixed model, which relates the $\tlY_i$ to the covariates 
\begin{equation}
\mbox{logit}\left\{\hbox{Pr}\left(\tlY_i=1\mid u_i,\blx_i\right)\right\}
= \psi_0(u_i)+\sum_{d=1}^px_{id}\psi_d(u_i) + \sum_{\ell=1}^Lr_\ell(i)\gamma_\ell,
\label{eqn:1}
\end{equation}
where $\mbox{logit}\{\cdot\}$ is the canonical logit link, and the regression coefficients $\psi_d(u)$'s are allowed to vary smoothly with $u$. Additionally, the clinic-specific random effects are represented by $r_\ell(\cdot)$, serving as the $\ell$th clinic indicator function; i.e., $r_\ell(i)=1$ indicates the association of the $i$th subject with the $\ell$th clinic-specific random effect $\gamma_\ell$, while $r_\ell(i)=0$ denotes otherwise, for $\ell=1,\ldots,L$. We assume the random effects $\gamma_\ell$'s follow $\mathcal{N}(0,\sigma^2)$ independently.
%are considered independent and identically distributed normal random variables with an unknown variance $\sigma^2$, i.e., $\gamma_\ell\sim\mathcal{N}(0,\sigma^2)$.

% is the main age effect, and $\alpha$

% . when $d=0$, $x_{i0}=1$ and the term 

% and for $d>1$, $\alpha_d+\beta_d(u_i)$ is the regression coefficient of the $d$th covariate.

% and fixed effects are denoted by parameters $\alpha_d$'s. In contrast, coefficients $\beta_d(\cdot)$'s are not constant but rather permitted to vary smoothly as functions of $u_i$. Additionally, the clinic-specific random effects are represented by $r_\ell(\cdot)$, serving as the $\ell$th clinic indicator function. Specifically, $r_\ell(i)=1$ indicates the association of the $i$th subject with the $\ell$th clinic-specific random effect $\gamma_\ell$, while $r_\ell(i)=0$ denotes otherwise, for $\ell=1,\ldots,L$. Throughout the article, the canonical logit link function $g^{-1}\{\cdot\}$ is assumed, and the random effects $\gamma_\ell$'s are considered independent and identically distributed normal random variables with an unknown variance $\sigma^2$, i.e., $\gamma_\ell\sim\mathcal{N}(0,\sigma^2)$.

In group testing, $\tlY_i$'s cannot be observed due to pooling and imperfect testing. To denote data from any group testing protocol, we let $J$ be the total number of pools that have been tested (for generality, if a specimen is tested individually, we view it as tested in a pool of size $1$). We use the index set $\mathcal{P}_j \subset \{1,\ldots, N\}$ to collect all individuals contributing to the $j$th pool,  for $j=1,\ldots, J$. Again, we permit $\mathcal{P}_j$ to be a singleton set, allowing for individual testing scenarios. Additionally, we require that $\mathcal{P}_j\neq\emptyset$ and  $\cup_j\cP_j=\{1,\dots, N\}$; i.e., each individual should be tested at least once either in pools or individually. 
%We use $\mathcal{P}_j\subset \{1,\ldots,N\}$ to index all the individuals contributing to the specimen tested by the $j$th test, for $j=1,\ldots, J$. We require $\mathcal{P}_j\neq\emptyset$ and  $\cup_j\cP_j=\{1,\dots, N\}$ (i.e., each individual has to be tested), and allow $\cP_j$ to be a singleton set (i.e., an individual can be tested individually).
%Various group testing algorithms have been proposed, such as the hierarchical protocol \citep{kotz1982errors} and the multistage protocol \citep{kennedy2004multistage}. In practice, once the algorithm is chosen, individuals undergo testing in a series of pools. While the testing algorithm can generate individual diagnoses, the final results exhibit high correlation due to inherent testing errors and pool structures. Consequently, it is essential to collectively consider all observed data from group testing.
%Regrettably, the true disease statuses $\tlY_i$'s are never directly observed due to the impact of pooling and imperfect testing. 
%, the observed group testing data structure can be intricate and
%, involving individuals in multiple overlapping pools and diverse testing protocols such as the hierarchical protocol \citep{kotz1982errors} and the multistage protocol \citep{kennedy2004multistage}. 
%To accommodate data arising from an arbitrary group testing algorithm, we let $J$ be the total number of tests used in the screening processes. 
 
The binary variable $\tlZ_j=\max_{i\in\mathcal{P}_j}\tlY_i=1$ (or $0$) indicates that the $j$th pool is \textit{truly} positive (or negative).  However, the $\tlZ_j$'s are latent due to imperfect testing. Instead, we observe the error-contaminated $Z_j$'s, where $Z_j=1$ (or $0$) if the $j$th pool \textit{tested} positively (or negatively). To evaluate the impact of imperfect testing, we denote the sensitivity and specificity of the assay used to test the $j$th pool by $S_{ej}=\hbox{Pr}(Z_j=1\mid \tlZ_j=1)$ and $S_{pj}=\hbox{Pr}(Z_j=0\mid \tlZ_j=0)$, respectively. Furthermore, it is commonly assumed in group testing literature \citep{mcmahan2017bayesian, joyner2020mixed, liu2021generalized} that conditional on $\tlZ_j$'s, $Z_j$'s are independent and do not depend on the covariates. Under these assumptions and based on \eqref{eqn:1}, the observed data likelihood can be written as
\begin{align}
\label{eqn:2}
\pi(\blZ\mid \tbY, \bmeta,\blS_e,\blS_p) 
=&
\sum_{{\scriptsize \tbY}\in\{0,1\}^N}\left[\prod\limits_{j=1}^J\left\{S_{ej}^{Z_j}\left(1-S_{ej}\right)^{1-Z_j}\right\}^{\tlZ_j}\left\{\left(1-S_{pj}\right)^{Z_j}S_{pj}^{1-Z_j}\right\}^{1-\tlZ_j}\right.\nonumber\\
&\times\left.\prod\limits_{i=1}^N g(\eta_i)^{\tlY_i}\{1-g(\eta_i)\}^{1-\tlY_i}\right],
\end{align}
where $\blZ$ collects all the $Z_j$'s, $\tbY$ collects all the $\tlY_i$'s, $\bmeta$ collects the $\eta_i=\psi_0(u_i)+\sum_{d=1}^px_{id}\psi_d(u_i) + \sum_{\ell=1}^Lr_\ell(i)\gamma_\ell$ for $i=1,\dots, N$, $\blS_e=(\blS_{e1},\ldots,\blS_{eJ})^\top$ and $\blS_p=(\blS_{p1},\ldots,\blS_{pJ})^\top$.
%$\eta_i=\sum_{d=0}^px_{id}\left\{\alpha_d+\beta_d(u_i)\right\} + \sum_{\ell=1}^Lr_\ell(i)\gamma_\ell$. For ease of notation, aggregations are employed as $\tbY=(\tlY_1,\tlY_2,\ldots,\tlY_N)^\top$, $\bmalpha=(\alpha_1,\alpha_2,\ldots,\alpha_p)^\top$, $\bmbeta=(\bmbeta_0^\top,\bmbeta_1^\top,\ldots,\bmbeta_p^\top)^\top$ with $\bmbeta_d=\{\beta_d(u_1), \beta_d(u_2), \ldots, \beta_d(u_N)\}^\top$, $\bmgamma=(\gamma_1,\gamma_2,\ldots,\gamma_L)^\top$, $\blS_e=(\blS_{e(1)},\ldots,\blS_{e(J)})^\top$, $\blS_p=(\blS_{p(1)},\ldots,\blS_{p(J)})^\top$. 
Notably, the summation in the right-hand side of \eqref{eqn:2} is numerically infeasible when $N$ is considerably large, whereas a two-step data augmentation procedure will be introduced to circumvent this issue and thus facilitates a computationally feasible posterior algorithm. 

\subsection{Variable selection preliminaries}\label{sec-variable-selection}

Our variable selection procedure aims to regularize the modeling fitting and mitigate possible overfitting. The procedure classifies each $\psi_d(\cdot)$ in \eqref{eqn:1} into one of three categories \citep[similar as in][]{reich2010bayesian, cai2013bayesian}. To be more specific, we rewrite 
%The framework we propose posits the concurrent consideration of two distinct effects associated with each covariate in $\blx_i$: a fixed effect and an age-varying effect. However, this configuration results in an overparameterized model. To induce sparsity in our model, a stochastic search variable selection technique is employed to retain many variables in the model with only a small subset exhibiting age-varying effects. To this end, we re-define $\eta_i$ in Equation \ref{eqn:2} as
\begin{align}\label{eqn:3}
\psi_d(u)=\delta_{1d}\{\alpha_d+\delta_{2d}\beta_d(u)\}
\end{align}
where $\delta_{1d}\in\{0,1\}$ and $\delta_{2d}\in\{0,1\}$ denote binary inclusion indicator variables for the main fixed effect $\alpha_d$ and the age-varying effect $\beta_d(\cdot)$ of the $d$th covariate, respectively. Under this construction, we consider three scenarios. (i) \textit{Insignificant}: $\psi_d(u)=0$; i.e., the $d$th covariate should be excluded from the model, or equivalently, $\delta_{1d}=0$. (ii) \textit{Significant but age-independent}: $\psi_d(u)=\alpha_d$; i.e., $\delta_{1d}=1$ and $\delta_{2d}=0$ and we estimate $\alpha_d$. (iii) \textit{Significant and age-varying}: $\psi_d(u)=\alpha_d+\beta_d(u)$; i.e., $\delta_{1d}=\delta_{2d}=1$ and we estimate both $\alpha_d$ and $\beta_d(u)$. Though we also write the main age effect $\psi_0(u)$ as in \eqref{eqn:3}, we assume it does vary in age and thus fix $\delta_{10}=\delta_{20}=1$, a mild assumption that can be easily relaxed if needed. %If $\beta_0(u)$ does not vary in age, then another covariate should be set as the varying index $u$.  
Furthermore, for model identification, we require $\sum_{i=1}^N\beta_d(u_i)=0$.%These assumptions are similar to the ones in for spatial varying coefficient models.
%In other words, we assume that a covariate lacks an age-varying effect in the absence of a fixed effect. 
%In line with \citep{reich2010bayesian, cai2013bayesian}, we assume the overall age effect (intercept) varies across different age groups, i.e., $\delta_{10}=\delta_{20}\equiv1$, accounting for missing covariates and/or reflecting age variations. Further, 

\section{Data augmentation and prior solicitation}\label{sec3}

\subsection{Data augmentation}\label{sec-data-augmentation}
The first step of our data augmentation method is to recast unobserved true statuses in $\bm\tlY=(\tlY_1,\ldots,\tlY_N)^\top$ as latent random variables. %, which modifies \eqref{eqn:2} to be
% \begin{align}
% \label{eqn:4}
% \pi(\blZ, \tbY\mid \bmeta,\blS_e,\blS_p) 
% =&
% \prod\limits_{j=1}^J\left\{S_{ej}^{Z_j}\left(1-S_{ej}\right)^{1-Z_j}\right\}^{\tlZ_j}\left\{\left(1-S_{pj}\right)^{Z_j}S_{pj}^{1-Z_j}\right\}^{1-\tlZ_j}\nonumber\\
% &\times \prod\limits_{i=1}^N g(\eta_i)^{\tlY_i}\{1-g(\eta_i)\}^{1-\tlY_i},
% \end{align}
% We will update $\tbY$ in the posterior sampling. This step helps us avoid the computational burden of summating $2^N$ terms in \eqref{eqn:2}.
The second step follows \cite{polson2013bayesian} to introduce the latent variables $\omega_i$'s, aggregated as $\bmomega=(\omega_1,\ldots,\omega_N)^\top$, for each of individuals, where $\omega_i$'s independently follow a Pólya-Gamma (PG) distribution with parameters $(1,0)$, denoted as $\hbox{PG}(1,0)$. Applying this two-step data augmentation procedure yields
\begin{align*}
%\label{eqn:5}
\pi(\blZ, \tbY,\bmomega \mid & \bmeta,\blS_e,\blS_p) 
\propto
\prod\limits_{j=1}^J\left\{S_{ej}^{Z_j}\left(1-S_{ej}\right)^{1-Z_j}\right\}^{\tlZ_j}\left\{\left(1-S_{pj}\right)^{Z_j}S_{pj}^{1-Z_j}\right\}^{1-\tlZ_j\nonumber}\\
&\times \exp\left\{-\frac{1}{2}\sum_{i=1}^N\omega_i(h_i-\eta_i)^2\right\}\prod\limits_{i=1}^N f(\omega_i\mid 1,0)\exp\left\{(\tlY_i-0.5)^2/(2\omega_i)\right\},
\end{align*}
where $h_i=(\tlY_i - 0.5)/\omega_i$ and $f(\cdot \mid 1,0)$ denotes the density function for $\hbox{PG}(1,0)$. 
These latent variables help us avoid the computational burden of summating $2^N$ terms in \eqref{eqn:2} and facilitate a simple and effective Bayesian inference for logistic regression models.

\subsection{Prior for the SSVS process}
The SSVS process is governed by the binary indicators, $\delta_{1d}$'s and $\delta_{2d}$'s. We set the prior of $(\delta_{1d},\delta_{2d})$ jointly by the following probability mass function,
%The prior for the remaining binary inclusion indicators constrains effects from varying across age groups unless the fixed effect is nonzero, for $d=1,\ldots,p$, as follows:
\begin{align*}
\pi(\delta_{1d},\delta_{2d}\mid\theta_{1d},\theta_{2d})=
\begin{cases}
1-\theta_{1d},\quad&(\delta_{1d},\delta_{2d})=(0,0),\\
\theta_{1d}(1-\theta_{2d}),\quad&(\delta_{1d},\delta_{2d})=(1,0),\\
\theta_{1d}\theta_{2d},\quad&(\delta_{1d},\delta_{2d})=(1,1),
\end{cases}
\end{align*}
% and denote this prior by $(\delta_{1d},\delta_{2d})\mid(\theta_{1d},\theta_{2d})\sim\text{Cat}(\pi_{\delta_{1d},\delta_{2d}},\mathcal{S})$ with $\mathcal{S}=\{(0,0),(1,0),(1,1)\}$. 
with the support $\mathcal{S}=\{(0,0),(1,0),(1,1)\}$. Herein, $\theta_{1d}=\hbox{Pr}(\delta_{1d}=1)$ is the probability of including the $d$th covariate, while $\theta_{2d}=\hbox{Pr}(\delta_{2d}=1\mid\delta_{1d}=1)$ is the probability of including an age-varying effect of the $d$th covariate given its inclusion in the model. We set the hyperparameters $\theta_{1d}\sim\hbox{Beta}(a_{\theta_{1d}},b_{\theta_{1d}})$ and $\theta_{2d}\sim\hbox{Beta}(a_{\theta_{2d}},b_{\theta_{2d}})$ independently for $d>0$. Our numerical studies have $a_{\theta_{1d}}=b_{\theta_{1d}}=a_{\theta_{2d}}=b_{\theta_{2d}}=1$, corresponding to the non-informative priors.

\subsection{Priors for regression coefficients and other parameters}\label{sec-gaussian-predictive-process}
%regression coefficients
For $\alpha_d$, we set its prior by $\alpha_d\sim\mathcal{N}(0,\xi_{\alpha_d})$. In practice, if there is no prior information about $\alpha_d$, one can set $\xi_{\alpha_d}$ to be large (e.g., we used $50$ in our analyses).
% with a large value of $\sigma^2_{\alpha_d}$ (e.g., $50$). 
For the age-varying coefficients,  $\beta_d(\cdot)$'s, we employ GPPs to estimate $\bmbeta_d=\{\beta_d(u_1),\dots, \beta_d(u_N)\}^\top$. Following \cite{banerjee2008gaussian}, GPP involves specifying $\tlK$ knots as $(\tlu_1,\ldots,\tlu_{\tlK})^\top$ and first focuses on $\tbbeta_d=\{\beta_d(\tlu_1),\ldots,\beta_d(\tlu_{\tlK})\}^\top$. The GPP assumes $\tbbeta_d\sim\mathcal{N}(\bm{0},\tbC_d)$ independently for $d=0,1,\ldots,p$. In the covariance matrix $\tbC_d=\tau_d^{-1}\tbR_d$, $\tau_d$ is the precision parameter, and $\tbR_d$ is the $\tlK\times\tlK$ correlation matrix, entries of which are specified using the Matérn function \citep{gneiting2010matern}. 
%The Matérn function \citep{gneiting2010matern} is used to specify the entries of $\tbR_d$. is denoted as $\rho_d(\tlu_{k_1},\tlu_{k_2}\mid \nu_d,\phi_d)$. 
Under GPPs, we have $\bmbeta_d$ related to $\tbbeta_d$ through $\bmbeta_d =\blE\blQ_d\tbbeta_d$ where $\blE$ is an $N\times K$ matrix and $\blQ_d$ is $K\times \tlK$. Explanation of $\blE$ and $\blQ_d$ and the Matérn function $\rho_d(\cdot,\cdot\mid\nu_d,\phi_d)$, which depends on $\nu_d$ for smoothness and $\phi_d$ for the decay rate, can be found in Web Appendix A.
%More details of the Matérn function, the matrices $\blE$ and $\blQ_d$ are included in Web Appendix A of the supplementary material.
%where $\blE$ is an $N\times K$ binary sparse matrix indicating the $K$ unique age values, and $\blQ_d$ is a $K\times \tlK$ dense matrix. We specifically choose the Matérn function described in \cite{gneiting2010matern}, where the $(k_1,k_2)$th entry of $\tbR_d$ is denoted as $\rho_d(\tlu_{k_1},\tlu_{k_2}\mid \nu_d,\phi_d)$. Derivations for $\blQ_d$ and the Matérn formula $\rho_d(\cdot,\cdot\mid\nu_d,\phi_d)$, which depends on $\nu_d$ for smoothness and $\phi_d$ for the decay rate, can be found in Web Appendix A of the supplementary material. 
In our analyses, we set $\tlK=100$, specify the prior for $\tau_d$ to be $\hbox{Gamma}(a_{\tau_d},b_{\tau_d})$ with $a_{\tau_d}=2$ and $b_{\tau_d}=1$, and set $\nu_d=2$ but learn $\phi_d$ from the data. For the variance of the random effect, we set $\sigma^2\sim\text{InverseGamma}(a_{\sigma^2},b_{\sigma^2})$ with $a_{\sigma^2}=2$ and $b_{\sigma^2}=1$. 

For the assay sensitivity and specificity, if $S_{ej}$ and $S_{pj}$ vary across $j=1,\dots, J$, estimation suffers from identifiability issues. Following \cite{mcmahan2017bayesian}, we assume there are $M$ assays or $M$ pairs $\{(S_{e(m)}, S_{p(m)}): m=1,\dots, M\}$ to be estimated, where $M$ is much less than $J$. For example, our analysis in Section \ref{sec5} considers $M=2$ with $S_{e(1)}$ and $S_{p(1)}$ being the sensitivity and specificity of the assay on pooled specimens, respectively, while $S_{e(2)}$ and $S_{p(2)}$ are the ones on individual specimens (see another example in Section \ref{sec6}). Under this assumption, we let $\cM_m=\{j: \cP_j \mbox{ tested by the assay with } (S_{e(m)}, S_{p(m)})\}$ for $m=1,\dots, M$. 
That is, $\cM_m$ identifies all the testing outcomes that can help us estimate $(S_{e(m)}, S_{p(m)})$. In our estimation, we set the priors of $S_{e(m)}$ and $S_{p(m)}$ to be $\text{Beta}(a_{S_{e(m)}},b_{S_{e(m)}})$ and $\text{Beta}(a_{S_{p(m)}},b_{S_{p(m)}})$, respectively, with $a_{S_{e(m)}}=a_{S_{p(m)}}=b_{S_{e(m)}}=b_{S_{p(m)}}=0.5$ adhere to Jeffreys priors \citep{gelman2009bayes}.
%do we need to emphasize that GPPs do provide a remedy for computational burdens incurred by GPs. 

\section{Posterior sampling}\label{sec4}
The proposed posterior sampling algorithm involves five steps. For ease of notation, let $\bmTheta=\{\tbY,\bmomega,\bmdelta_1,\bmdelta_2,\bmtheta_1,\bmtheta_2,\bmlambda,\bmtau,\bmphi,\bmgamma,\sigma^2,\blS_e,\blS_p\}$ be the complete set of latent variables and model parameters, where $\bmdelta_1$, $\bmdelta_2$, $\bmlambda$, $\bmtau$ and $\bmphi$ are the collections of all $\delta_{1d}$'s, $\delta_{2d}$'s, $\theta_{1d}$'s, $\theta_{2d}$'s, $\bmlambda_d=(\alpha_d,\tbbeta_d^\top)^\top$'s, $\tau_d$'s and $\phi_d$'s, respectively. In the following, we use the notation $\bmTheta_{-\bm{\vartheta}}$ to denote the subset of $\bmTheta$ excluding the element $\bm{\vartheta}$. Note that we only outline the five steps below. A detailed version is included in the Web Appendix B.
%denote the set excluding $\bm{\vartheta}$ as $\bmTheta_{-\bm{\vartheta}}$,  Each $\bmlambda_d=(\alpha_d,\tbbeta_d^\top)^\top$ specifies regression coefficients for the $d$th covariate.

\textit{Step 1: Sample latent random variables, $\tbY$ and $\bmomega$}. To sample $\tlY_i$, we observe $\tlY_i\mid \bmTheta_{-\tlY_{i}}\sim \hbox{Bernoulli}\left\{p_{i1}^*/\left(p_{i0}^*+p_{i1}^*\right)\right\}$ with $p_{i0}^*$ and $p_{i1}^*$ provided in \textit{Step 1(a)} in the Web Appendix B. Following the properties of the PG distribution \citep{polson2013bayesian}, we can quickly sample $\omega_i\mid \bmTheta_{-\omega_i} \sim \hbox{PG}(1,\eta_i)$. 

\textit{Step 2: Sample binary inclusion indicators $(\bmdelta_1,\bmdelta_2)$ and $(\bmtheta_1,\bmtheta_2)$.} For $d>0$, sampling $(\delta_{1d},\delta_{2d})$ directly from its full conditional posterior distribution $\pi(\delta_{1d},\delta_{2d}\mid\bmTheta_{-(\delta_{1d},\delta_{2d})})$ can lead to an absorbing state in the Markov chain and ruin the posterior sampling. Instead, we draw $(\delta_{1d},\delta_{2d})$ from its the marginal posterior conditional distribution by firstly integrating $\bmlambda_d$ out. Routine algebra shows that
\begin{align}\label{eqn:marginal}
\pi\left\{\delta_{1d},\delta_{2d}\mid \bmTheta_{-(\delta_{1d},\delta_{2d},\bmlambda_d)}\right\}
&\propto
|\bmSigma_d|^{1/2}\exp\left(\frac{1}{2}\bmmu_d^\top\bmSigma_d\bmmu_d\right)\times \pi(\delta_{1d},\delta_{2d}\mid \theta_{1d},\theta_{2d}),
\end{align}
where $\bmmu_d$ and $\bmSigma_d$ are explicitly presented in \textit{Step 2(a)} of the Web Appendix B. From \eqref{eqn:marginal}, we can sample each $(\delta_{1d},\delta_{2d})\in\mathcal{S}$ from their posterior probabilities accordingly. To sample $\theta_{1d}$ and $\theta_{2d}$, one can obtain $\theta_{1d}\mid\delta_{1d}\sim\hbox{Beta}(a_{\delta_{1d}}+\delta_{1d}, b_{\delta_{1d}}+1-\delta_{1d})$ and $\theta_{2d}\mid\delta_{2d}\sim\hbox{Beta}(a_{\delta_{2d}}+\delta_{2d}, b_{\delta_{2d}}+1-\delta_{2d})$. 

\textit{Step 3: Sample regression coefficients.} Given the values of $(\delta_{1d},\delta_{2d})$, one can sample $\alpha_d$ and $\tbbeta_d$ accordingly. To be more specific, if $(\delta_{1d},\delta_{2d})=(0,0)$, we set $\bmlambda_d=\bm{0}$ and hence $\alpha_d=0$ and $\tbbeta_d=\bm{0}$; if $(\delta_{1d},\delta_{2d})=(1,0)$, we set $\tbbeta_d=\bm{0}$ and sample $\alpha_d\mid\bmTheta_{-\bmlambda_d}\sim\mathcal{N}(\mu_{\alpha_d}^*,\xi_{\alpha_d}^*)$; otherwise, one can observe $\bmlambda_d\mid\bmTheta_{-\bmlambda_d}\sim\mathcal{N}(\bmmu_d,\bmSigma_d)$ from \eqref{eqn:marginal}. Derivations for $\mu_{\alpha_d}^*$, $\xi_{\alpha_d}^*$ are provided in \textit{Step 3(a)} in the Web Appendix B. To sample $\tau_d$, one can obtain $\tau_d\mid\bmTheta_{-\tau_d}\sim\hbox{Gamma}(a_{\tau_d}+\tlK/2,b_{\tau_d}+\tbbeta_d^\top\tbR_d^{-1}\tbbeta_d/2)$. For $\phi_d$, we follow the Metropolis-Hastings algorithm (Algorithm S.1) in Web Appendix B.

\textit{Step 4: Sample random effects.} To sample $\gamma_{\ell}$, one can derive that $\gamma_\ell\mid\bmTheta_{-\gamma_\ell}\sim\mathcal{N}(\mu_{\gamma_\ell}, \sigma^2_{\gamma_\ell})$ given $\gamma_\ell\sim\mathcal{N}(0,\sigma^2)$. Derivations for $\mu_{\gamma_\ell}$, $\sigma^2_{\gamma_\ell}$ are provided in \textit{Step 4(a)} in the Web Appendix B. For $\sigma^2$, we update $\sigma^2 \mid \bmTheta_{-\sigma^2} \sim\hbox{InverseGamma}(a_{\sigma^2}+L/2, b_{\sigma^2}+\sum_{\ell=1}^L\gamma_\ell^2/2)$.

\textit{Step 5: Sample assay accuracy probabilities.} Given the beta priors $S_{e(m)}\sim \hbox{Beta}(a_{S_{e(m)}},b_{S_{e(m)}})$ and $S_{p(m)}\sim \hbox{Beta}(a_{S_{p(m)}},b_{S_{p(m)}})$, the conditional posterior distributions are $S_{e(m)}\mid\blZ,\tbY\sim\hbox{Beta}(a^*_{S_{e(m)}},b^*_{S_{e(m)}})$ and $S_{p(m)}\mid\blZ,\tbY\sim\hbox{Beta}(a^*_{S_{p(m)}},b^*_{S_{p(m)}})$, where $a^*_{S_{e(m)}}=a_{S_{e(m)}}+\sum_{j\in\mathcal{M}_m}Z_j\tlZ_j$, $b^*_{S_{e(m)}}=b_{S_{e(m)}}+\sum_{j\in\mathcal{M}_m}(1-Z_j)\tlZ_j$, $a^*_{S_{p(m)}}=a_{S_{p(m)}}+\sum_{j\in\mathcal{M}_m}(1-Z_j)(1-\tlZ_j)$, and $b^*_{S_{p(m)}}=b_{S_{p(m)}}+\sum_{j\in\mathcal{M}_m}Z_j(1-\tlZ_j)$. 

%To accommodate different assays (e.g., screening assay and confirmatory assay) and justify the impact of pool size on assay accuracy, we adopt the index set $\mathcal{M}_m=\{j:\text{the $m$th assay tests the $j$th pool}\}$ for $m=1, \ldots, M$, following \cite{mcmahan2017bayesian}, to track pools tested by the $m$th assay, preserving the generality. That is, the sensitivity and specificity of the $m$th assay satisfy that $S_{ej}=S_{e(m)}$ and $S_{pj}=S_{p(m)}$ for all $j\in\mathcal{M}_m$, which necessitates sampling only $S_{e(m)}$ and $S_{p(m)}$. Posterior updates for $S_{e(m)}\mid\bm{Z},\bm{Y}\sim\hbox{Beta}(a^{*}_{S_{e(m)}},b^{*}_{S_{e(m)}})$ and $S_{p(m)}\mid \bm{Z},\bm{Y}\sim \hbox{Beta}(a^{*}_{S_{p(m)}},b^{*}_{S_{p(m)}})$ are given, with parameters specified in \textit{Step 5} of Web Appendix B. 

\section{Simulation}\label{sec5}

\subsection{Data generation}
To evaluate our method, we design a simulation study that replicates key aspects of the SHL group testing data. We create a clinic network with $L=64$ sites and generate infection statuses for $N$ individuals, who are uniformly distributed among these clinics. The true infection statuses $\tlY_i$'s are generated following the model below:
\begin{align*}
\mbox{logit}\left\{\hbox{Pr}\left(\tlY_i=1\mid u_i,\blx_{i}\right)\right\}=\psi_0(u_i)+\sum_{d=1}^6x_{id}\left\{\alpha_d+\beta_d(u_i)\right\} + \sum_{\ell=1}^Lr_{\ell}(i)\gamma_\ell.
\end{align*}
Herein, $\gamma_\ell\sim\mathcal{N}(0,\sigma^2)$ with $\sigma=0.5$, for $\ell=1,\ldots, L$, are independent random effects. To mimic the covariates pattern in the SHL dataset, we simulate the age variable $u_i\sim\hbox{Uniform}(-3,3)$ with rounding to the nearest hundredth and set $\blx_i=(x_{i1},x_{i2},x_{i3},x_{i4},x_{i5},x_{i6})^\top$, where $x_{i1}\sim\mathcal{N}(0,1)$, and $x_{i2}$ to $x_{i6}$ each following $\hbox{Bernoulli}(0.5)$. The true value of $\bmalpha$ is $(\alpha_0,\alpha_1,\ldots,\alpha_6)^\top=(-3.5, -1.0, 0.5, -0.5, 0.5, 0, 0)^\top$. To evaluate our variable selection, we set $\beta_1(u)$, $\beta_3(u)$, $\beta_5(u)$, $\beta_6(u)$ to be zero, and model $\beta_0(u)$, $\beta_2(u)$, $\beta_4(u)$ as smooth age-varying functions. We consider two model sets for these age-varying coefficients to cover various non-linear patterns:
\begin{align*}
\hbox{M1}&:\begin{cases}
\beta_0(u)=\sin(\pi u/3)\\
\beta_2(u)=u^3/8\\
\beta_4(u)=-u^2/4 + 3/4
\end{cases}&
\hbox{M2}&:\begin{cases}
\beta_0(u)=-0.5\exp\{-\sin(u)\}+0.64\\
\beta_2(u)=0.3x^2+\sin^2(u/3)-0.9\\
\beta_4(u)=\Phi(u)-0.5
\end{cases}
\end{align*}
 where $\Phi(\cdot)$ is the cumulative distribution function of $\mathcal{N}(0,1)$. To be consistent with the real data application, the carefully designed parameters above maintain an approximately $9\%$ disease prevalence. In summary, our generating model emphasizes estimating and selecting two significant but age-independent effects ($\alpha_1$, $\alpha_3$), three age-varying effects \{$\beta_0(\cdot)$, $\beta_2(\cdot)$, $\beta_4(\cdot)$\}, and two insignificant effects ($\alpha_5$, $\alpha_6$). 

We considered the two-stage array testing (AT) algorithm proposed in \cite{phatarfod1994use} in addition to the Dorfman testing (DT) employed at SHL. The DT algorithm randomly assigns individuals to non-overlapping pools of size $c$, while the AT algorithm randomly assigns them to size $c\times c$ arrays. In AT, the initial stage involves mixing specimens in rows and columns to create row and column pools, which are subsequently tested. In the second stage, specimens with a higher likelihood of being positive (e.g., individuals at the intersection of positive rows and columns) undergo individual retesting \citep[more details are referred to][]{kim2007comparison}. 

%We follow two group testing protocols to generate the testing response $\blZ$: Dorfman Testing (DT) from \cite{dorfman1943detection} and Array Testing (AT) from \cite{phatarfod1994use}. The former one is used in SHL, while the latter one is also a two-stage group testing algorithm. In DT, non-overlapping master pools of sizes $c$ are created from individual specimens and subjected to testing in the initial stage. Positive master pools are subsequently deciphered through individual retesting in the second stage. In AT, we first assign individuals to non-overlapping $c\times c$ arrays. The first stage tests all the row and column pools formed by mixing specimens in the rows and columns, respectively. At the second stage, specimens likely to be positive (such as individuals at the intersection of positive rows and columns) are retested individually \citep[see more details in][]{kim2007comparison}.  

%To generate the testing responses, $\blZ$, we randomly assign the $N$ individuals to pools or arrays of size $c$ or $c\times c$, respectively, for $c\in \{5,10\}$ by grouping across and within sites. In this way, specimens within a single pool or array exhibit distinct spatial random effects. 
When generating the testing outcomes $Z_j$, we consider two assays (i.e., $M=2$). %to accommodate the influence of the dilution effect, which relates to the case that testing accuracies on pools might be different than the ones on individuals). 
The pool responses under each protocol are simulated using the first assay with $S_{e(1)}=0.95$ and $S_{p(1)}=0.98$. Individual tests/retests are simulated using the second assay with $S_{e(2)}=0.98$ and $S_{p(2)}=0.99$. In short, for any $\cP_j$, $S_{ej}=I(|\cP_j|>1)S_{e(1)}+I(|\cP_j|=1)S_{e(2)}$ and $S_{pj}=I(|\cP_j|>1)S_{p(1)}+I(|\cP_j|=1)S_{p(2)}$. The testing response on $\cP_j$ is generated by $Z_j \mid \tlZ_j \sim \hbox{Bernoulli}\{S_{ej}\tlZ_j+(1-S_{pj})(1-\tlZ_j)\}$, where $\tlZ_j=\max_{i\in\mathcal{P}_j}\tlY_i$.

We have simulated data for each combination of the sample size ($N=3000$ or $N=5000$), the model setting (M1 or M2), the pool size ($c=5$ or $c=10$), and the testing protocol (DT or AT). For comparative reasons, individual testing (IT) is also implemented. The entire process has been repeated $500$ times to assess the performance of our methodology comprehensively. In our estimation, our proposed algorithm draws $15000$ iterations, with every $5$th iteration retained after a burn-in of $5000$ samples. These numbers were chosen to ensure consistent mixing and convergence, as verified by trace plots. In each replication, we estimate $\psi_d(\cdot)$'s, $\sigma^2$, $S_{e(m)}$'s, and $S_{p(m)}$'s using the respective posterior medians.

\subsection{Results}\label{sec:simu-results}
\begin{figure}[!htbp]
\centerline{\includegraphics[width=1.0\textwidth]{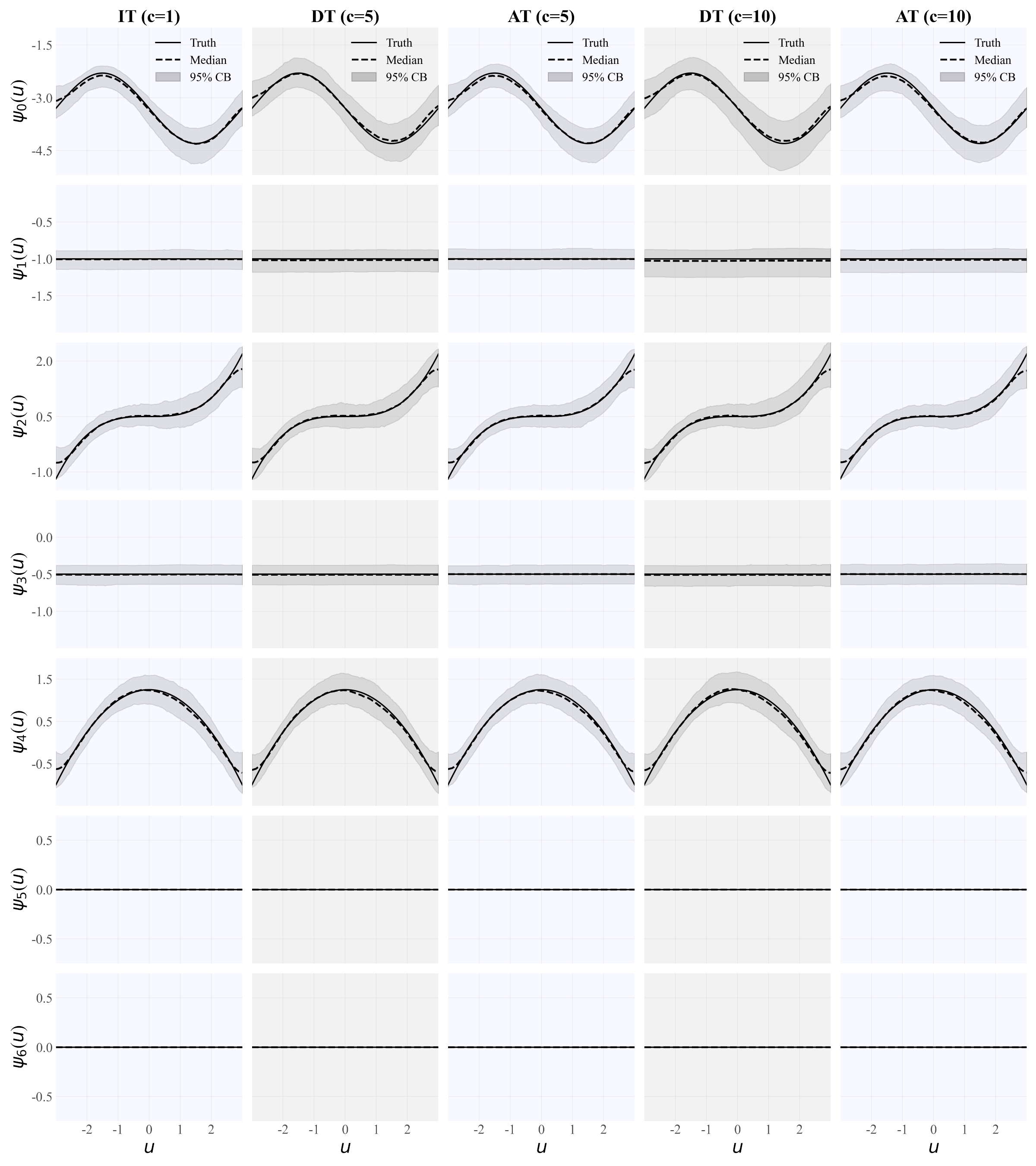}}
\caption{Simulation results for model M1 with sample size $N=5000$, $c\in\{5,10\}$ under both DT and AT protocols. Each subfigure features a solid curve for the true value of the varying coefficient (Truth), a dashed curve for the pointwise median of the $500$ posterior median estimates (Median), and a gray-shaded area for the equal-tailed 95\% credible band (95\% CB) with the lower and upper bounds being the 2.5\% and 97.5\% pointwise quantiles of the $500$ posterior median estimates. Results for IT are also presented as a reference.}
\label{fig:m1N5000}
\end{figure}

In this section, we present the outcomes for M1 with $N=5000$, while results for other scenarios are provided in Web Appendix C. Figure \ref{fig:m1N5000} summarizes our $500$ posterior median estimates of the regression coefficient functions, $\psi_d(\cdot)$ for $d=0,1,\dots, 6$. When $\psi_d(\cdot)\neq 0$, our method well estimates these coefficients, whether they vary with $u$ ($d=0,2,4$) or remain constant ($d=1,3$). Notably, the pointwise median of the $500$ $\psi_d(u)$ estimates (depicted with dashed lines) closely aligns with the true values (represented by solid lines), exhibiting minimal bias where present, while the pointwise equal-tailed 95\% credible bands accurately envelop the true curves across the entire support. It's reassuring to observe that the width of credible bands remains almost constant. %particularly for instances where $\psi_d(u)=\alpha_d$ remains constant across $u$. %Interestingly, the comparisons between DT/AT and IT reveal little loss of efficiency despite that DT/AT can save testing costs greately.
When $\psi_d(\cdot)=0$ for $d=5$ or $6$, it is promising to see that the median line and 95\% credible bands are unified to be exactly $0$, indicating our method's success in identifying the insignificant covariates. Further evidence of the efficacy of our variable selection method is demonstrated in Table \ref{tab:m1N5000}.

%The results in Figure \ref{fig:m1N5000} show close agreement between pointwise median estimates and the true functions, with pointwise credible regions covering them completely. These results not only validate the efficiency of our proposed algorithm in identifying significant age-independent and age-dependent effects for $\beta_d(\cdot)$, $d>0$, but also demonstrate its robustness in capturing non-linear patterns. Furthermore, the DT and AT protocols demonstrate minimal variability and comparable estimation performance to IT, requiring fewer tests, and this holds with larger pool sizes. There is minimal distinction between M1 and M2, and larger sample sizes lead to improved estimation and reduced variability. Through our extensive simulation study, the proposed method has exhibited robustness in identifying significant age-independent, significant age-dependent, and insignificant effects, while effectively capturing non-linear patterns. 

Table \ref{tab:m1N5000} first summarizes the performance of our SSVS process. For evaluation, we consider the inclusion probability (IP) of the $d$th covariate; i.e., $\text{Pr}(\delta_{1d}=1)$. Since $\delta_{2d}$ is binary, this probability can be written as the summation of two:
(i) IPF, the inclusion probability of the fixed effect  $\alpha_d$ but not the varying effect $\beta_d(u)$, which is $\Pr(\delta_{1d}=1,\delta_{2d}=0)$; (ii) IPV, the inclusion probability of the varying effect $\beta_d(u)$, which is $\Pr(\delta_{1d}=1,\delta_{2d}=1)$. We note that IP is the summation of IPF and IPV. %The last probability we consider is the CPI, $\Pr(\delta_{2d}=1|\delta_{1d}=1)$, which is the conditional probability of including the varying effect $\beta_d(u)$ given $x_d$ being significant. 
All these probabilities can be estimated using the respective posterior means in each simulation. We report the averages of these estimates from our 500 replications in Table \ref{tab:m1N5000} for $d>0$.

\begin{sidewaystable}[!htbp]
\caption{Simulation results for model set M1 with $N=5000$ and $c\in\{5,10\}$ under both DT and AT. Summary statistics include Bias (empirical bias), SSD (standard deviation of the 500 posterior median estimates), ESE (average of the 500 estimates of the posterior standard deviation), CP95 (empirical coverage probability of equal-tail 95\% credible intervals), AVGtest (average number of tests used), and savings (percentage reduction in average number of tests compared to IT). For the performance of the SSVS, inclusion probabilities (IP, IPF, and IPV) are displayed (see Section \ref{sec:simu-results} for their definitions).\label{tab:m1N5000}} 
\small{
\begin{center}
\begin{tabular}{llrcrrcrr}
\toprule
\midrule
\multicolumn{3}{c}{\bfseries }&\multicolumn{1}{c}{\bfseries }&\multicolumn{2}{c}{\bfseries \bm{$c=5$}}&\multicolumn{1}{c}{\bfseries }&\multicolumn{2}{c}{\bfseries \bm{$c=10$}}\tabularnewline
\cline{5-6} \cline{8-9}\addlinespace[0.1cm]
\multicolumn{1}{l}{Parameter}&\multicolumn{1}{l}{Summary}&\multicolumn{1}{c}{IT}&\multicolumn{1}{c}{}&\multicolumn{1}{c}{DT}&\multicolumn{1}{c}{AT}&\multicolumn{1}{c}{}&\multicolumn{1}{c}{DT}&\multicolumn{1}{c}{AT}\tabularnewline
\midrule
% $\alpha_0=-3.3$
% &Bias(CP95)&-0.041(0.956)&&0.038(0.914)&-0.029(0.928)&&0.020(0.814)&-0.027(0.910)\tabularnewline
% $\delta_{10}=1$
% &SSD(ESE)&0.129(0.139)&&0.178(0.151)&0.136(0.134)&&0.248(0.172)&0.167(0.150)\tabularnewline
% $\delta_{20}=1$
% &IPF/IPV/CPI&0.000/1.000/1.000&&0.000/1.000/1.000&0.000/1.000/1.000&&0.000/1.000/1.000&0.000/1.000/1.000\tabularnewline
% \addlinespace[0.2cm]
% $\alpha_5=0.0$
% &Bias(CP95)&0.000(1.000)&&0.000(1.000)&0.000(1.000)&&0.000(1.000)&0.000(1.000)\tabularnewline
% $\delta_{15}=0$
% &SSD(ESE)&0.007(0.009)&&0.000(0.010)&0.006(0.010)&&0.007(0.009)&0.000(0.009)\tabularnewline
% $\delta_{25}=0$
% &IPF/IPV/CPI&0.007/0.000/0.021&&0.007/0.000/0.014&0.007/0.000/0.015&&0.007/0.000/0.020&0.007/0.000/0.020\tabularnewline
% \addlinespace[0.2cm]
% $\alpha_6=0.0$
% &Bias(CP95)&0.000(1.000)&&0.000(1.000)&0.000(1.000)&&0.000(1.000)&0.000(1.000)\tabularnewline
% $\delta_{16}=0$
% &SSD(ESE)&0.010(0.010)&&0.008(0.010)&0.009(0.010)&&0.000(0.009)&0.003(0.009)\tabularnewline
% $\delta_{26}=0$
% &IPF/IPV/CPI&0.007/0.000/0.020&&0.007/0.000/0.019&0.008/0.000/0.019&&0.007/0.000/0.019&0.007/0.000/0.018\tabularnewline
%$\alpha_5=0.0$ &Bias(CP95)&0.000(1.000)&&0.000(1.000)&0.000(1.000)&&0.000(1.000)&0.000(1.000)\tabularnewline
%$\delta_{15}=0$ &SSD(ESE)&0.007(0.009)&&0.000(0.010)&0.006(0.010)&&0.007(0.009)&0.000(0.009)\tabularnewline
$\psi_5(u)=0$ %&IPF/IPV &0.007/0.000 &&0.007/0.000 &0.007/0.000 &&0.007/0.000 &0.007/0.000 
&IP &0.007 &&0.007 &0.007 &&0.007 &0.007 \tabularnewline%&IPF/IPV/CPI&0.007/0.000/0.021&&0.007/0.000/0.014&0.007/0.000/0.015&&0.007/0.000/0.020&0.007/0.000/0.020\tabularnewline
%$\alpha_6=0.0$ &Bias(CP95)&0.000(1.000)&&0.000(1.000)&0.000(1.000)&&0.000(1.000)&0.000(1.000)\tabularnewline
%$\delta_{16}=0$ &SSD(ESE)&0.010(0.010)&&0.008(0.010)&0.009(0.010)&&0.000(0.009)&0.003(0.009)\tabularnewline
$\psi_6(u)=0$  %&IPF/IPV &0.007/0.000 &&0.007/0.000 &0.008/0.000 &&0.007/0.000 &0.007/0.000 \tabularnewline 
&IP &0.007 &&0.007 &0.008 &&0.007 &0.007 \tabularnewline %&IPF/IPV/CPI&0.007/0.000/0.020&&0.007/0.000/0.019&0.008/0.000/0.019&&0.007/0.000/0.019&0.007/0.000/0.018\tabularnewline
\addlinespace[0.1cm]
\cline{2-9}
\addlinespace[0.1cm]
$\psi_1(u)=-1.0$ &IPF/IPV&0.994/0.005 &&0.993/0.007 &0.993/0.007 &&0.993/0.007 &0.993/0.007 \tabularnewline
$\psi_3(u)=-0.5$ &IPF/IPV&0.993/0.007 &&0.993/0.007 &0.994/0.006 &&0.987/0.014 &0.989/0.011 \tabularnewline
\addlinespace[0.1cm]
\cline{2-9}
\addlinespace[0.1cm]
%$\alpha_2=0.5$ &Bias(CP95)&0.006(0.966)&&0.007(0.948)&0.005(0.944)&&0.007(0.954)&-0.003(0.948)\tabularnewline
%$\delta_{12}=1$ &SSD(ESE)&0.076(0.082)&&0.077(0.078)&0.077(0.077)&&0.085(0.084)&0.077(0.081)\tabularnewline
%$\delta_{22}=1$
$\delta_{12}=1,\delta_{22}=1$ &IPF/IPV &0.000/1.000 &&0.000/1.000 &0.000/1.000 &&0.000/1.000 &0.000/1.000 \tabularnewline
%$\alpha_4=0.5$ &Bias(CP95)&0.000(0.958)&&-0.002(0.962)&-0.003(0.952)&&0.005(0.954)&-0.003(0.966)\tabularnewline
%$\delta_{14}=1$ &SSD(ESE)&0.075(0.078)&&0.075(0.075)&0.072(0.074)&&0.083(0.081)&0.074(0.078)\tabularnewline
%$\delta_{24}=1$
$\delta_{14}=1,\delta_{24}=1$  &IPF/IPV &0.000/1.000 &&0.000/1.000 &0.000/1.000 &&0.000/1.000 &0.000/1.000 \tabularnewline
\midrule
\multirow{2}{*}{$\psi_1(u)=\alpha_1=-1.0$}
&Bias(CP95)&-0.008(0.962)&&-0.015(0.940)&-0.004(0.942)&&-0.034(0.912)&-0.013(0.932)\tabularnewline
%\delta_{11}=1$
&SSD(ESE)&0.066(0.069)&&0.072(0.072)&0.068(0.067)&&0.095(0.083)&0.077(0.074)\tabularnewline
\addlinespace[0.1cm]
%$\delta_{21}=0$
\multirow{2}{*}{$\psi_3(u)=\alpha_3=-0.5$}
&Bias(CP95)&-0.007(0.942)&&-0.008(0.936)&-0.002(0.938)&&-0.012(0.938)&-0.001(0.938)\tabularnewline
%$\delta_{13}=1$
&SSD(ESE)&0.064(0.063)&&0.067(0.063)&0.067(0.062)&&0.072(0.068)&0.066(0.065)\tabularnewline
\addlinespace[0.1cm]
\multirow{2}{*}{$\sigma=0.5$}&Bias(CP95)&0.047(0.950)&&0.044(0.952)&0.039(0.964)&&0.059(0.924)&0.048(0.960)\tabularnewline
&SSD(ESE)&0.062(0.079)&&0.062(0.080)&0.058(0.078)&&0.067(0.084)&0.062(0.081)\tabularnewline
\midrule
\multirow{2}{*}{$S_{e(1)}=0.95$}&Bias(CP95)&&&-0.032(0.902)&0.000(0.954)&&-0.038(0.914)&0.002(0.964)\tabularnewline
&SSD(ESE)&&&0.081(0.053)&0.024(0.021)&&0.084(0.062)&0.041(0.034)\tabularnewline
\addlinespace[0.1cm]
\multirow{2}{*}{$S_{e(2)}=0.98$}&Bias(CP95)&&&-0.008(0.974)&-0.001(0.988)&&-0.019(0.972)&-0.006(0.996)\tabularnewline
&SSD(ESE)&&&0.021(0.024)&0.016(0.017)&&0.050(0.046)&0.021(0.033)\tabularnewline
\addlinespace[0.1cm]
\multirow{2}{*}{$S_{p(1)}=0.98$}&Bias(CP95)&&&0.002(0.990)&0.000(0.920)&&-0.014(0.992)&-0.010(0.990)\tabularnewline
&SSD(ESE)&&&0.011(0.014)&0.012(0.011)&&0.032(0.052)&0.027(0.033)\tabularnewline
\addlinespace[0.1cm]
\multirow{2}{*}{$S_{p(2)}=0.99$}&Bias(CP95)&&&0.000(0.966)&-0.003(0.974)&&-0.003(0.922)&-0.003(0.966)\tabularnewline
&SSD(ESE)&&&0.008(0.007)&0.012(0.013)&&0.011(0.009)&0.012(0.011)\tabularnewline
\midrule
Cost &AVGtest(savings)&5000.00(00.00\%)&&2943.15(41.14\%)&2971.33(40.57\%)&&3567.84(28.64\%)&2943.73(41.12\%)\tabularnewline
% &AVGtime(mins)&273.787(mins)&&283.214(mins)&266.855(mins)&&176.97(mins)&168.631(mins)\tabularnewline
\midrule
\bottomrule
\end{tabular}\end{center}}
\end{sidewaystable}

When $\psi_d(\cdot)=0$, the $d$th covariate is insignificant to the model, meaning $\Pr(\delta_{1d}=1)=0$. Hence, IP should be close to $0$, as evident in Table \ref{tab:m1N5000} for $d\in\{5,6\}$. Conversely, when $\psi_d(\cdot)\neq 0$, we have two cases: (i) \textit{significant but age-independent} as $\psi_d(u)=\alpha_d$, where IPF should be close to $1$ while IPV approaches $0$; (ii) \textit{significant and age-varying} as $\psi_d(u)=\alpha_d+\beta_d(u)$, where IPF should be close to $0$ while IPV approaches $1$. In Table \ref{tab:m1N5000}, these patterns are evident for $d\in\{1,3\}$ and $d\in\{2,4\}$, underscoring the effectiveness of our SSVS approach.

For $d\in\{1,3\}$, it's notable that $\psi_d(u)$ reduces to a single value $\alpha_d$ . In addition to $\sigma$, $S_{e(m)}$'s, and $S_{p(m)}$'s, Table \ref{tab:m1N5000} summarizes our estimates of these parameters across all considered testing protocols. However, we lack results for estimating $S_{e(m)}$'s and $S_{p(m)}$'s under IT since they are not identifiable in this scenario. Examination of these summary statistics reveals minimal to no bias in the estimates, close agreement between SSDs and ESEs, and mostly nominal CP95s. These observations further underscore the inferential capabilities of our method.

Finally, we'd like to discuss the testing protocols. The last row of Table \ref{tab:m1N5000} indicates that both DT and AT achieve significant cost reductions (28\%--41\%) compared to IT, without sacrificing estimation accuracy as both DT and AT protocols at $c=5$ or $c=10$ exhibit minimal variability and comparable estimation performance to IT. These trends are also evident in the results presented in Web Appendix C. In conclusion, our comprehensive simulation study demonstrates that the proposed group testing method is capable of identifying significant age-independent, significant age-varying, and insignificant covariates while achieving comparable performance to IT of adeptly capturing nonlinear patterns and accurately estimating other model parameters but with great cost-savings.

\section{Application}\label{sec6}

We now apply our method to the SHL chlamydia dataset. This data was obtained from screening $N=13862$ females over $L=64$ clinics throughout Iowa in 2014. The screening involved both individual testing and group testing. The individual testing was conducted on all $4316$ urine specimens and $416$ swab specimens. The remaining $9130$ swabs were tested in pools following the DT algorithm. In the group testing part, it tested $2286$ swab master pools ($2273$ pools of size $4$, $12$ of size $3$, and $1$ of size $2$) with additional retests in resolving positive pools. 

In addition to all the testing outcomes, the dataset also contains covariates for each participant, denoted by $u_i$ and $\blx_i=(x_{i1},x_{i2},x_{i3},x_{i4},x_{i5},x_{i6},x_{i7},x_{i8})^\top$ for $i=1,\dots, N$. Specifically, $u_i$ denotes the $i$th female's age and all $x_{id}$'s are binary: $x_{i1}=1$ for Caucasian participants (and 0 otherwise), $x_{i2}=1$ if a new sexual partner was reported in the last 90 days (and 0 otherwise), $x_{i3}=1$ if multiple partners were reported in the last 90 days (and 0 otherwise), $x_{i4}=1$ if the subject had contact with a partner reporting any STD in the previous year (and 0 otherwise), $x_{i5}=1$ if the patient experienced symptoms of infection such as painful urination/intercourse (and 0 otherwise), $x_{i6}=1$ the female experienced symptoms associated with a friable cervix such as painful intercourse or frequent spotting (and 0 otherwise), $x_{i7}=1$ if the subject experienced the cervicitis, the inflammation of the cervix (and 0 otherwise), and $x_{i8}=1$ if the patient has been diagnosed with the pelvic inflammatory disease (PID), an infection of the female reproductive organs (and 0 otherwise).

%We note that all the urine specimens were tested separately, while the swabs were through the DT procedure.

To explore whether the association between the $p=8$ covariates (especially race) and chlamydia infection risk varies with age, we fit the following model,
\begin{align*}
%\label{eqn:app}
\hbox{logit}\left\{\hbox{Pr}\left(\tlY_i=1\mid u_i,\blx_i\right)\right\}
=&\psi_0(u_i)+\sum_{d=1}^{8}x_{id}\delta_{1d}\{\alpha_d+\delta_{2d}\beta_d(u_i)\}+\sum_{\ell=1}^Lr_{\ell}(i)\gamma_\ell,
\end{align*}
for $i=1,2,\ldots, N$, where $Y_i$ is the  $i$th female's true (latent) infection status, $\psi_0(\cdot)=\alpha_0+\beta_0(\cdot)$ captures the main age-varying effect (or the intercept term), and for $d>0$, $\delta_{1d}\{\alpha_d+\delta_{2d}\beta_d(\cdot)\}=\psi_d(\cdot)$ depicts the $d$th covariate's age-varying effect with $(\delta_{1d},\delta_{2d})$ classifying $\psi_d(\cdot)$ as either significant age-independent, significant age-varying, or insignificant. The random effects $\gamma_\ell$'s follow $\mathcal{N}(0,\sigma^2)$ independently and account for possible spatial heterogeneity across the $64$ clinic sites, and $r_\ell(i)=1$ if $i$th individual specimen was collected at the $\ell$th clinic. 

%We also posit $\delta_{10}=\delta_{20}=1$ to accommodate missing variables and/or
%account for age variations as assumed in the simulation. 

In our model fitting, different sensitivity and specificity parameters are used to account for the testing errors on different specimens. More specifically, we denote by $S_{e(1)}$ and $S_{p(1)}$, respectively, the sensitivity and specificity of the test on the individual swab specimens. For the individual test on the urine specimens, they are denoted by $S_{e(2)}$ and $S_{p(2)}$, and for the test on pooled swab specimens, we use $S_{e(3)}$ and $S_{p(3)}$. This formulation accounts for nuances in assay performance, considering the differences between urine and swab specimens and the distinction between individual and pooled tests. In our analysis, all prior specifications are aligned with those detailed in Section \ref{sec3}. For comparison, we also fit the model  without our SSVS process; i.e,
\begin{align*}
%\label{eqn:app2}
\hbox{logit}\left\{\hbox{Pr}\left(\tlY_i=1\mid u_i,\blx_i\right)\right\}
=&\psi_0(u_i)+\sum_{d=1}^{8}x_{id}\psi_d(u_i)+\sum_{\ell=1}^Lr_{\ell}(i)\gamma_\ell.
\end{align*}
In the posterior sampling of both models, we discarded the first 5000 draws for MCMC convergence and retained every 50th sample from the subsequent 20000 iterations. 

%%%%%%%%%%%%%%%%%%%%%%%%% conclusion

The posterior mean estimate of the standard deviation, $\sigma$, of the random effect is  $0.426$ with a 95\% highest posterior density (HPD) credible interval of $(0.313, 0.545)$. This indicates that the random effect is strongly significant, providing clear evidence of heterogeneity across clinics statewide. This result reinforces earlier findings from \cite{chen2009group} and \cite{joyner2020mixed}. Regarding the six assay accuracy probabilities, our findings align with those reported in \cite{mcmahan2017bayesian} and \cite{liu2021generalized} and are thus presented in Web Table S.4 in Web Appendix D.

%the results are similar to what has been found in \cite{liu2021generalized}, therefore, summarized in Web Table S.4 in the Web Appendix C of the Supplementary Material. 

\begin{figure}[!htbp]
\centerline{\includegraphics[width=1.0\textwidth]{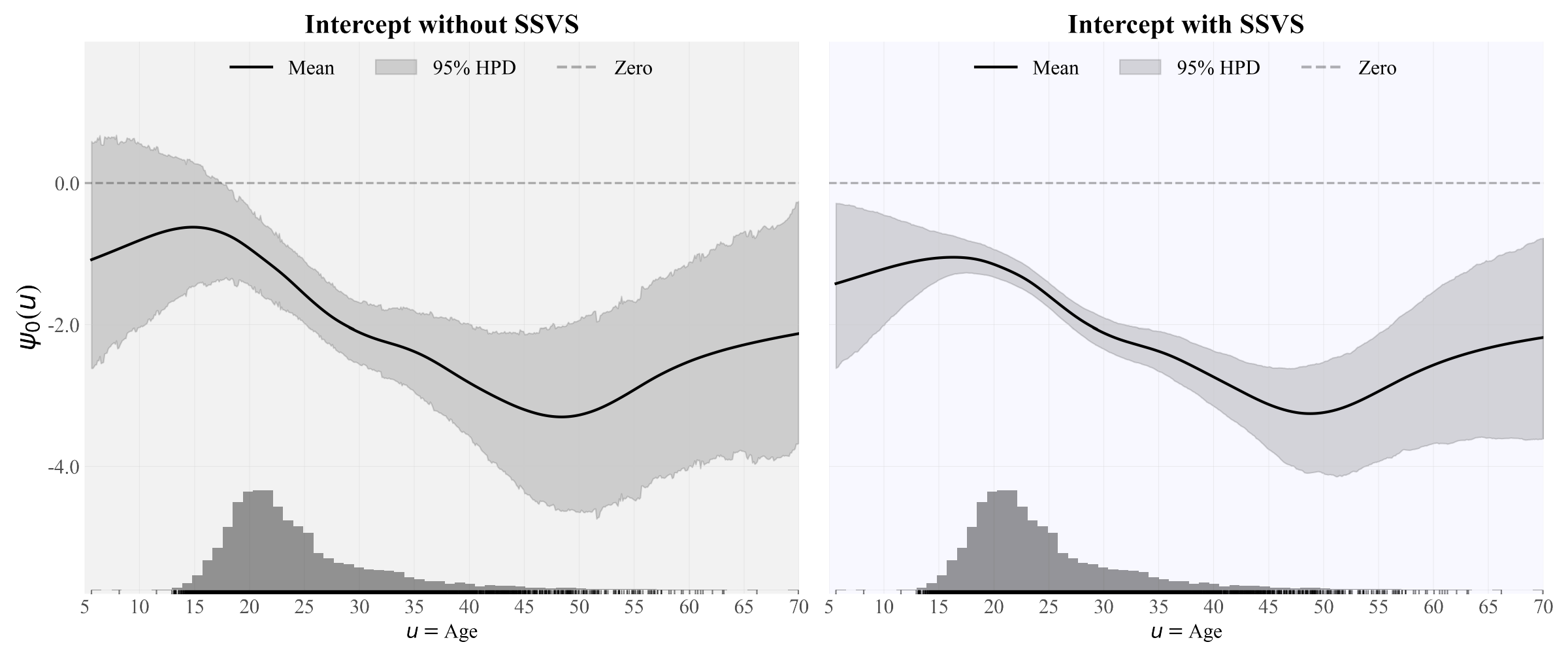}}
\caption{Iowa chlamydia data: the main age effect. Each subfigure displays the pointwise posterior mean (Mean) estimates and the pointwise 95\% HPD credible band (95\% HPD). The zero reference line is dashed. All the females' ages are plotted in dots and summarized in a (rescaled) histogram on the age-axis.}
\label{fig:compare}
\end{figure}

We now turn our attention to the
regression coefficients. We start with the estimates of the main age effect $\psi_0(u)$ that are summarized in Figure \ref{fig:compare}. 
The left and right panels present estimates of $\psi_0(u)$ without and with SSVS, respectively. Therein, solid lines depict the pointwise posterior mean estimates of the curve, while shaded areas represent the pointwise 95\% HPD credible bands. The age-axis has the ages of all female individuals as dots, accompanied by a (rescaled) histogram illustrating the distribution of these ages. Upon comparison, it becomes evident that although our SSVS regularizes the other covariates, it still benefits the estimate of $\psi_0(u)$. Notably, the credible band on the right is much narrower than that on the left, highlighting the non-linear trend in the estimates of $\psi_0(u)$. It's worth mentioning that this non-linear pattern is similar to the one identified by the generalized partially linear additive model in \cite{liu2021generalized}, which highlights a peak in infection risk around the age of 18, alongside a noticeable rise in risk for individuals aged 50 and above. We acknowledge that the non-linear patterns observed in the age groups 5--13 and 45--70 offer less informative insights compared to the age range 13--45, likely due to the limited availability of data near those boundaries, a phenomenon commonly known as the boundary effect in nonparametric estimation. The peak around 18 is more informative and aligns closely with CDC screening recommendations for chlamydia in sexually active females aged 24 or younger \citep{cdc2021}.

\begin{figure}[!htbp]
\centerline{\includegraphics[width=1.0\textwidth]{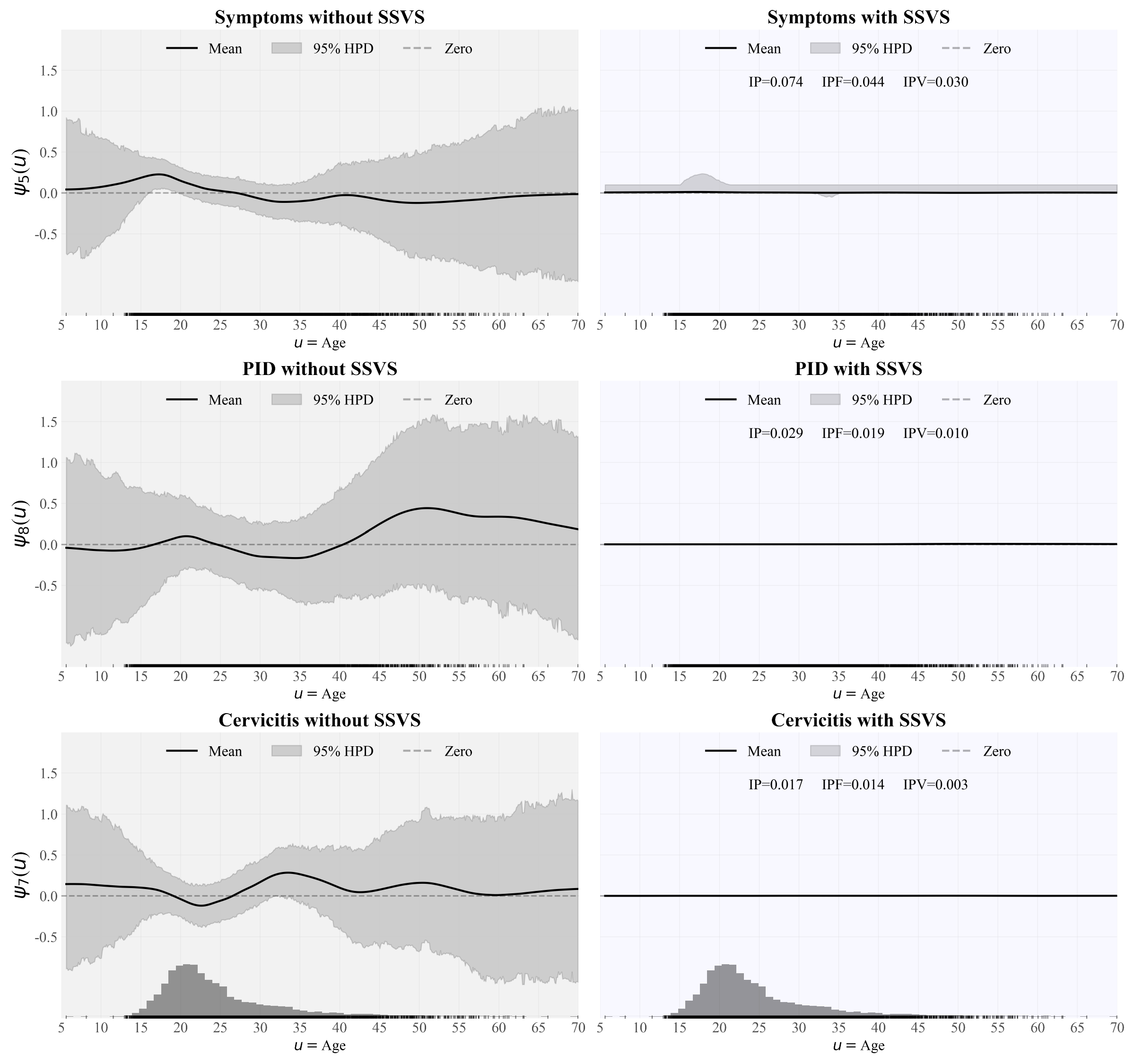}}
\caption{Iowa chlamydia data: insignificant covariates. Each subfigure displays the pointwise posterior mean (Mean) estimates and the pointwise 95\% HPD credible band (95\% HPD). The zero reference line is dashed. All the females' ages are plotted in dots and summarized in a (rescaled) histogram on the bottom age-axis. With SSVS, the estimated IP, IPF, and IPV are also provided.}
\label{fig:ssvs_trivial}
\end{figure}

For covariates, $x_1,\dots,x_8$, we categorize them into three groups based on the SSVS outputs. Those with an estimated $\mbox{IP}\leq 0.1$ are considered insignificant covariates. We choose $0.1$ as the threshold for two reasons: (1) it's a commonly used significance level in standard statistical hypothesis testing, and (2) we would rather be a bit conservative than overlook any significant covariates. Based on this criteria, the insignificant covariates identified are $x_5$, $x_7$, and $x_8$, with IPs of $0.074$, $0.017$, and $0.029$, respectively. Figure \ref{fig:ssvs_trivial} provides a summary of these estimates. In the case of $\psi_7(\cdot)$ and $\psi_8(\cdot)$, the figures on the left (without SSVS) present non-linear estimates but with a wide 95\% HPD credible band covering zero across all ages. Unsurprisingly, SSVS regularization precisely sets them to zero. In the corresponding figures, both the posterior mean estimate and the 95\% HPD credible intervals have been unified to the zero horizontal line, providing compelling evidence that PID or cervicitis does not significantly associate with chlamydia infection risk. Similarly, $x_5$ (symptoms) is also deemed insignificant, indicating that the presence of symptoms does not impact the risk of chlamydia infection. This aligns with expectations, as chlamydia infections frequently occur without symptoms yet can still attack a woman's reproductive system. Moreover, the SSVS credible band of $\psi_5(u)$ becomes slightly larger between ages 15 and 22 compared to other age groups, likely due to this age range being the most vulnerable period to chlamydia.

\begin{figure}[!htbp]
\centerline{\includegraphics[width=1.0\textwidth]{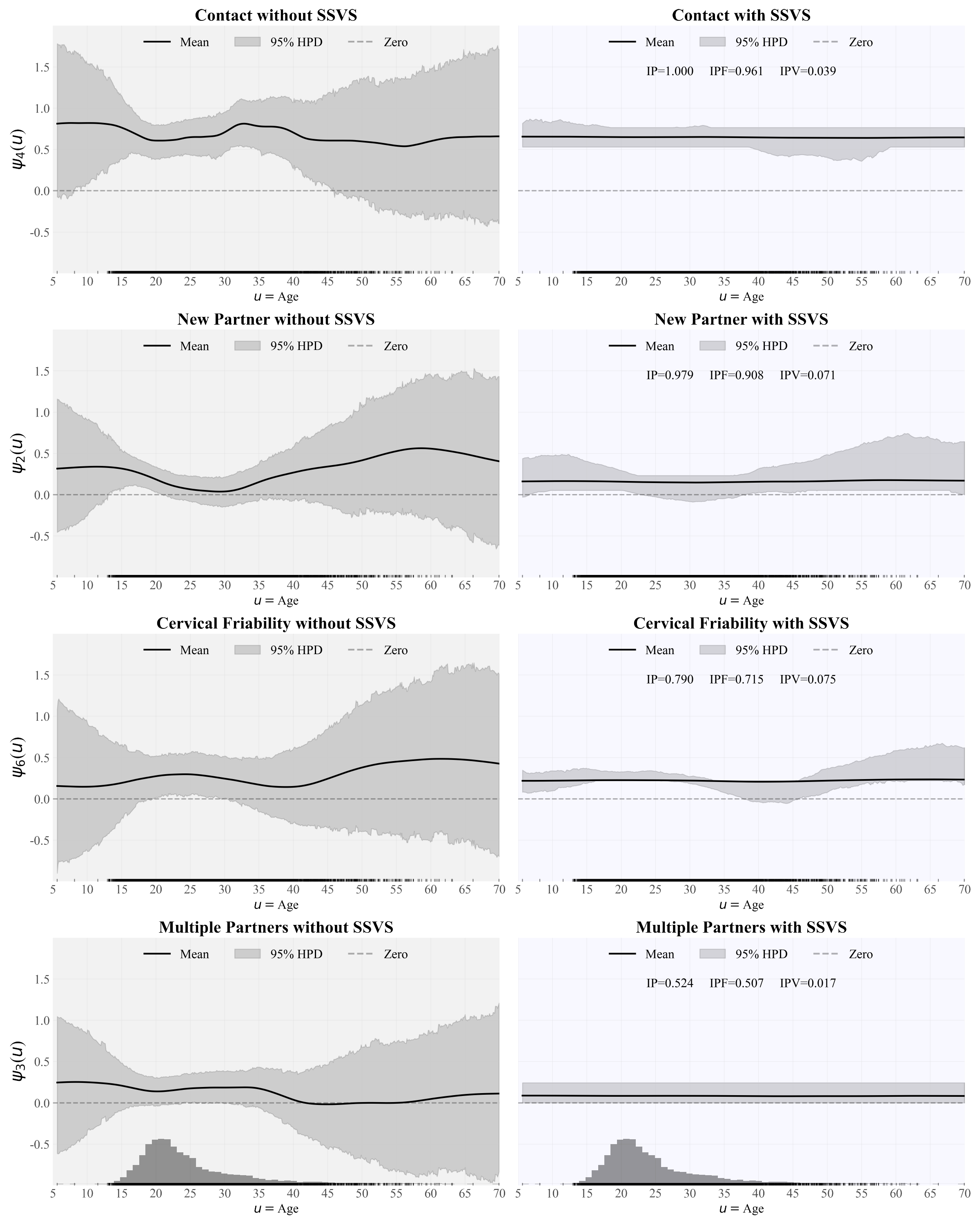}}
\caption{Iowa chlamydia data: significant and age-independent covariates. Each subfigure displays the pointwise posterior mean (Mean) estimates and the pointwise 95\% HPD credible band (95\% HPD). The zero reference line is dashed. All the females' ages are plotted in dots and summarized in a (rescaled) histogram on the bottom age-axis. With SSVS, the estimated IP, IPF, and IPV are also provided.}
\label{fig:ssvs_fix}
\end{figure}

Those covariates with estimated $\mbox{IP}>0.1$ and $\mbox{IPV}\leq 0.1$ are considered significant age-independent. This category includes $x_2$, $x_3$, $x_4$, $x_6$, with $\mbox{IP}=0.979, 0.524, 1.000, 0.979$ and $\mbox{IPV}=0.071, 0.017, 0.039, 0.075$, respectively. The corresponding estimates are summarized in Figure \ref{fig:ssvs_fix}. Again, employing SSVS results in substantially narrower credible bands compared to analyses without SSVS. All posterior mean estimates, with a majority of their credible bands, remain above zero, suggesting that chlamydia risk escalates with factors such as having contact with STDs ($x_2=1$), acquiring a new sexual partner ($x_3=1$), engaging in multiple partnerships ($x_4=1$), or experiencing a friable cervix ($x_4=1$). This finding aligns with known epidemiological trends of chlamydia \citep{lefevre2014screening}. %That is, chlamydia risk escalates with recent sexual activities, including acquiring new or multiple partners within six months, and is significantly heightened by having contact with STDs. 

\begin{figure}[!htbp]
\centerline{\includegraphics[width=1.0\textwidth]{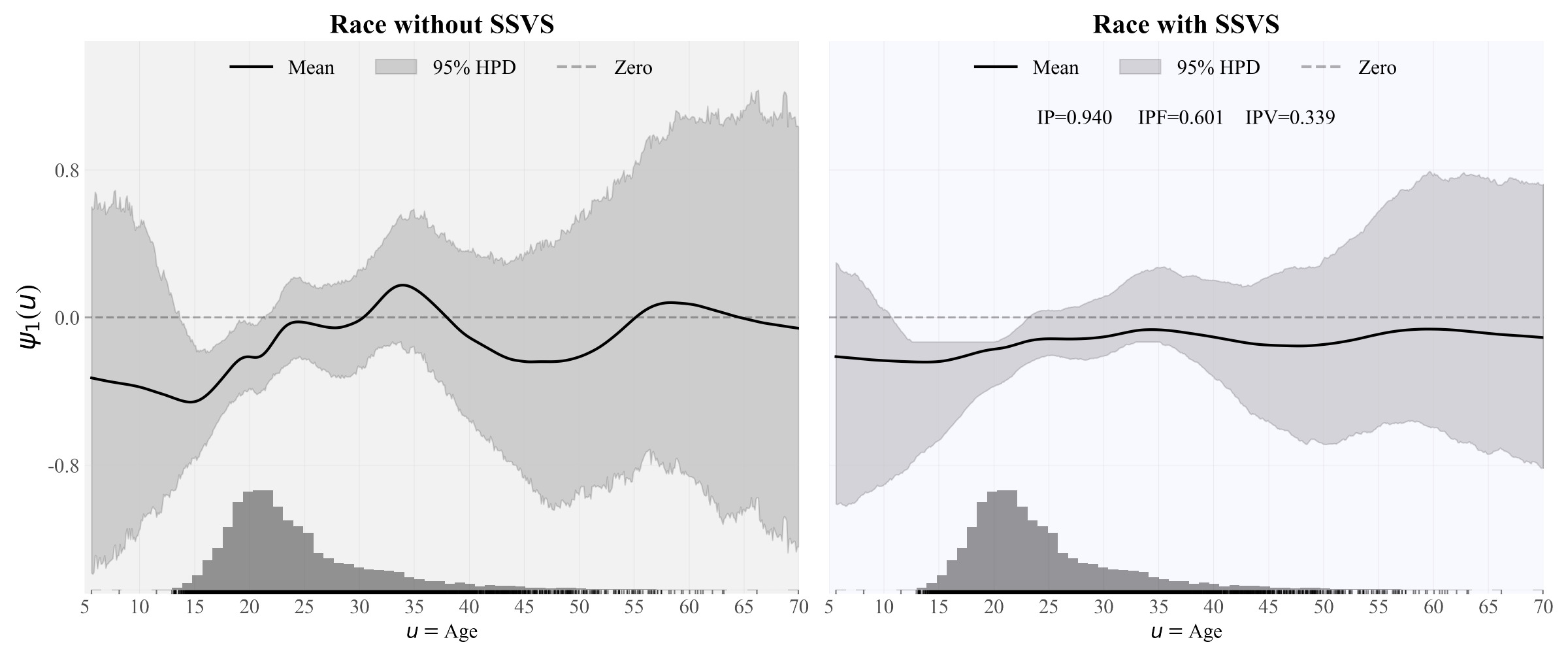}}
\caption{Iowa chlamydia data: the significant and age-varying covariate. Each subfigure displays the pointwise posterior mean (Mean) estimates and the pointwise 95\% HPD credible band (95\% HPD). The zero reference line is dashed. All the females' ages are plotted in dots and summarized in a (rescaled) histogram on the bottom age-axis. With SSVS, the estimated IP, IPF, and IPV are also provided.}
\label{fig:ssvs_varying}
\end{figure}

Covariates exhibiting estimated $\mbox{IP}>0.1$ and $\mbox{IPV}>0.1$ are considered significant age-varying. In this analysis, only race ($x_1$) falls into this category, with $\mbox{IP}=0.94$ and $\mbox{IPV}=0.339$. With the implementation of SSVS, the posterior mean estimate of $\psi_1(u)$ displays reduced fluctuation along with a narrower credible interval in Figure \ref{fig:ssvs_varying} as expected. A particularly intriguing discovery emerges within the age bracket of 11 to 23. Within this range, there are 7571 females, constituting approximately 54.6\% of the sample size, and the SSVS estimate of $\psi_1(u)$, along with its credible band, is consistently negative throughout this span. This suggests that individuals who are non-Caucasian ($x_1=0$) between the ages of 11 and 23 exhibit greater vulnerability compared to people who are Caucasian ($x_1=1$), with this discrepancy gradually diminishing after age $25$. This finding closely aligns with results reported in \cite{chambers2018racial}. We believe that this finding not only highlights the crucial need for utilizing varying-coefficient models when analyzing group testing data but also prompts policymakers to consider the age-varying racial disparity in developing interventions and prevention strategies for better chlamydia control and management.

\section{Discussion}\label{sec7}

This article introduces a Bayesian varying-coefficient model tailored for group testing data, enabling regression coefficients to vary with a chosen covariate, estimating unknown test accuracies, and providing an option for integrating spatial random effects. The model utilizes GPP prior distribution to estimate these varying coefficients. To ensure informative inference, we employ the SSVS process for regularization and categorize covariates into three groups. Through careful data augmentation, we develop a computationally efficient posterior sampling algorithm. Simulation results consistently demonstrate the method's effectiveness in estimating regression parameters and performing variable selection. These techniques are applied to analyze SHL group testing data from Iowa's chlamydia screening, revealing an intriguing age-varying pattern in the racial disparity of the disease.

As previously discussed in \cite{liu2021generalized}, it's important to acknowledge that the chlamydia dataset analyzed here wasn't obtained from a random sample of Iowa females. Rather, it may be considered representative of the ``highest-risk" residents. This aspect should be taken into account when interpreting the outcomes of our data analysis. However, this characteristic neither undermines the significance of our study's contribution nor restricts the applicability of our method to a random sample if one is available.

Future directions within this research line involve exploring more flexible approaches to modeling the varying coefficients. For instance, one could expand upon Bayesian additive regression trees (BART) \citep{chipman2010bart} to estimate these coefficients. BART offers greater flexibility compared to GPP priors, allowing for the capture of discontinuously varying functions and automatic determination of covariate importance for variable selection. Another possibility is to use Dirichlet processes (DP) \citep{gelfand2005bayesian,cai2013bayesian} for age-varying coefficients, alleviating biases associated with the normality assumption of age-varying coefficients inherent in GPP. Furthermore, one could explore the development of a multivariate varying coefficient model \citep{reich2010bayesian,zhu2012multivariate}, in which, each coefficient is a function of both age and race, facilitating more flexibility in understanding racial disparities. Regarding testing accuracies, another avenue for investigation could be to address the dilution effect in testing errors, where testing sensitivity might decrease when pooling a positive specimen with multiple negative ones. Previous work by \cite{wang2015general} may provide valuable insights in this regard. Lastly, there is potential to extend this work by developing joint varying-coefficient modeling methods that incorporate testing responses from multiplex assays. These assays, which test for multiple diseases simultaneously, have gained popularity among many public health labs.

\backmatter

\section*{Funding}

This research was supported by National Institutes of Health grants R03-AI135614 and R01-AI121351.

\section*{Supplementary Materials}

Web Appendices A--D, referenced in Sections~\ref{sec3}--\ref{sec6}, are available with the submission. The R code and documentation for simulation are publicly available at the GitHub repository: \texttt{https://github.com/yizenglistat/rvcm4gt}.

\bibliographystyle{biom}
\bibliography{ms}

\label{lastpage}

\end{document}